\documentstyle[prd,aps,epsfig,preprint,tighten]{revtex}
\begin{document}
\preprint{                                                BARI-TH/284-97}
\draft
\title{                  The solar neutrino problem              	\\
             after three hundred days of data at SuperKamiokande 	}
\author{         G.~L.~Fogli, E.~Lisi, and D.~Montanino			}
\address{   Dipartimento di Fisica and Sezione INFN di Bari, 		\\
                  Via Amendola 173, I-70126 Bari, Italy			}
\maketitle
\begin{abstract}
%...........................................................................
We present an updated analysis of the solar neutrino problem in terms of 
both Mikheyev-Smirnov-Wolfenstein (MSW) and vacuum neutrino oscillations, 
with the inclusion of the preliminary data collected by the SuperKamiokande
experiment during 306.3 days of operation. In particular, the observed 
energy spectrum of the recoil electrons from $^8$B neutrino scattering 
is discussed in detail and is used to constrain the mass-mixing parameter 
space. It is shown that: 1) the small mixing MSW solution is preferred 
over the large mixing one; 2) the vacuum oscillation solutions are 
strongly constrained by the energy spectrum measurement; and 3) the 
detection of a possible semiannual modulation of the $^8$B $\nu$ flux due 
to vacuum oscillations should require at least one more year of operation 
of SuperKamiokande.  
%...........................................................................
\end{abstract}

%%%%%%%%%%%%%%%%%%%%%%%%%%%%%%%%%%%%%%%%%%%%%%%%%%%%%%%%%%%%%%%%%%%%%%%%%%%%
\section{Introduction}
%%%%%%%%%%%%%%%%%%%%%%%%%%%%%%%%%%%%%%%%%%%%%%%%%%%%%%%%%%%%%%%%%%%%%%%%%%%%

	The solar neutrino problem \cite{Ba89}, namely, the deficit of the 
neutrino rates measured by the four pioneering solar neutrino experiments,
Homestake \cite{Da94}, Kamiokande \cite{Hi90}, SAGE \cite{Ab94}, and 
GALLEX \cite{An94}, as compared to the standard solar model predictions
\cite{BP95}, represents one of the most convincing indications for new 
physics beyond the standard electroweak theory. On the one hand, the 
confidence in the performances of the four detectors has increased 
continuously, as a result of many careful experimental cross-checks and 
calibrations \cite{Cl95,Fu96,Ga96,An95}. On the other hand, the reliability
of the most refined solar evolution models (including light element 
diffusion) has been corroborated by the impressive agreement with a 
growing amount of helioseismological data \cite{Ba97,Ri97}. Therefore, 
new neutrino physics (such as neutrino oscillations) appears to be a likely 
solution of the solar $\nu$ deficit.

	However, the deficit of the neutrino event rate is a solar model 
dependent quantity. It is highly desirable, instead, to measure observables
whose interpretation is not related to a prior knowledge of the absolute 
neutrino flux, such as shape deviations of energy spectra, or relative 
variations of the event rates during the time of the year. Such 
measurements have been  pioneered by the Kamiokande experiment \cite{Fu96}, 
but with statistics too low to provide definitive indications. This 
experimental program is now being pursued with much higher statistics by 
the real-time, water-Cherenkov SuperKamiokande experiment \cite{SupK}.

	The preliminary results after the first 306.3 days of operation of
SuperKamiokande \cite{To97,Na97,In97} are already sufficiently accurate 
to provide interesting new insights into the solar neutrino problem, 
although further data are required to draw decisive conclusions.
Therefore, we think it useful to present an updated analysis of the solar 
neutrino problem, including the most recent results from the four 
pioneering experiments \cite{La97,Fu96,SAGE,Cr97}  and from the 
SuperKamiokande experiment after 306.3 days \cite{To97,Na97,In97}, 
that have become publicly available during the 1997 summer conferences. 
In particular, we discuss in detail the information provided by the energy 
spectrum  of recoil electrons observed at SuperKamiokande, which is one 
of the most important solar model independent observables. The available 
experimental information is interpreted in the light of two-flavor
neutrino oscillations, both in vacuum \cite{JuSo} and in matter, according 
to the Mikheyev-Smirnov-Wolfenstein (MSW) mechanism \cite{MSWs}.

	The paper is organized as follows. In Sec.~II we analyze the solar 
neutrino problem using the available solar model dependent information 
(namely, the absolute neutrino rates). In Sec.~III we analyze the solar 
model independent information, and in particular the electron energy 
spectrum measured by SuperKamiokande, in order to constrain the oscillation
interpretation both in matter and in vacuum. We draw our conclusions
in Sec.~IV. Some technical aspects of the energy spectrum analysis are
elucidated in the Appendix.

%%%%%%%%%%%%%%%%%%%%%%%%%%%%%%%%%%%%%%%%%%%%%%%%%%%%%%%%%%%%%%%%%%%%%%%%%%%%
\section{Solar model dependent information}
%%%%%%%%%%%%%%%%%%%%%%%%%%%%%%%%%%%%%%%%%%%%%%%%%%%%%%%%%%%%%%%%%%%%%%%%%%%%

	In this section we report the most recent measurements of the 
solar neutrino rates, and compare them to the expectations of the 1995 
Bahcall-Pinsonneault (BP95) standard solar model \cite{BP95}. The observed 
deficit is interpreted in terms of MSW and vacuum two-family oscillations 
in the parameter space spanned by the neutrino squared mass difference 
$\delta m^2$ and by the mixing angle $\theta$.

%%%%%%%%%%%%%%%%%%%%%%%%%%%%%%%%%%%%%%%%%%%%%%%%%%%%%%%%%%%%%%%%%%%%%%%%%%%%
\subsection{The solar neutrino deficit}

	Table~I reports the neutrino event rates measured by the four 
pioneering solar neutrino experiments Homestake \cite{La97}, Kamiokande 
\cite{Fu96}, SAGE \cite{SAGE}, and GALLEX \cite{Cr97}, together with the 
$^8$B neutrino flux measurement after 306.3 days at SuperKamiokande 
\cite{To97,Na97,In97}. The observed rates appear to be significantly 
smaller than the theoretical expectations of the BP95 model.

	Figure~1 shows graphically the information reported in Table~I, 
with the further inclusion of the correlations of theoretical 
uncertainties, which have been calculated as in \cite{Erro}. In Fig.~1 
the GALLEX and SAGE data have been combined (in quadrature) in a 
single (Gallium) result.  The same has been done for the Kamiokande and 
SuperKamiokande data (K+SuperK). In the combination, asymmetric errors 
have been conservatively symmetrized to the largest one. The 99\% C.L.\ 
experimental and theoretical ellipses appear to be distinctly separated 
in all the planes charted by any two experiments. Notice that the 
(strongly correlated) theoretical errors are rather large as compared to 
the experimental errors. Therefore, as far as the total neutrino rates 
are concerned, further  reductions of the  experimental uncertainties 
are not expected to change significantly the current picture of the 
solar $\nu$ deficit, while improvements in the theoretical predictions 
would have a greater impact.

%%%%%%%%%%%%%%%%%%%%%%%%%%%%%%%%%%%%%%%%%%%%%%%%%%%%%%%%%%%%%%%%%%%%%%%%%%%%
\subsection{MSW and vacuum oscillation fits}

	As is well known, a viable explanation of the solar neutrino 
deficit is represented by matter-enhanced (MSW) neutrino oscillations 
(see, e.g.,\cite{Fo94,Fo96,Ha97}) or, alternatively, by vacuum neutrino 
oscillations (see, e.g., \cite{Kr92,Ha97}). Assuming for simplicity  
oscillations between two families, the analysis involves only two 
parameters, the neutrino squared mass difference $\delta m^2$ and the 
mixing angle $\theta$.

	Figure~2 shows the results of our $\chi^2$-analysis of the data 
reported in Table~I, assuming MSW oscillations. Notice that this fit 
includes only the $\nu_e$ deficit data, and not the solar model independent
information provided by the energy spectra or the night-day asymmetry
(that will be considered separately in Sec.~III). The usual small and large
mixing angle solutions appear to be slightly more constrained than in
previous fits (see, e.g., \cite{Fo96}), mainly as a result of the new
SuperKamiokande data. The absolute minimum  of $\chi^2$ 
$(\chi^2_{\rm min}=1.0)$ is reached at 
$(\delta m^2,\,\sin^2 2\theta)=({9.6\times10^{-6}\rm\ eV}^2,\,4.8\times
10^{-3})$; the secondary minimum  $(\chi^{2}_{\rm sec}=2.45)$ is reached at 
$(\delta m^2,\,\sin^2 2\theta)=({1.5\times10^{-5}\rm\ eV}^2,\,0.58)$.
The 90, 95, and 99\% C.L.\ allowed regions coincide with the contours
at $\chi^2-\chi^2_{\rm min}=4.61$, 5.99, and 9.21, respectively.

	Figure~3 shows the analogous results for the vacuum oscillation 
analysis. We find the minimum value $\chi^2_{\rm min}=4.0$ at 
$(\delta m^2,\,\sin^2 2\theta)=({8.2\times10^{-11}\rm\ eV}^2,\,0.87)$.
A comparison of $\chi^2_{\rm min}$ seems to indicate a preference
of the MSW oscillation scenario over the vacuum oscillation one. However, 
it should be noted that, in vacuum oscillation fits, the value
of $\chi^2_{\rm min}$  is very sensitive to small shifts in the central 
values of the experimental results and thus it might change significantly
with new data (while it turns out to be more stable in MSW fits).

	A few technical remarks are in order. In our oscillation analyses, 
the GALLEX and SAGE data have been combined in quadrature. However, the 
Kamiokande and SuperKamiokande data have been fitted separately, since 
these two experiments have different thresholds and energy resolutions. 
In particular, according to \cite{In97}, we use for SuperKamiokande a 
threshold of 6.5 MeV for the total electron energy $E_e$, and a Gaussian 
energy resolution function with a $1\sigma$ width of $\pm 15\%$ at $T=10$ 
MeV $(T=E_e-m_e$), scaling as $1/\sqrt{T}$ at different electron kinetic 
energies. The input neutrino fluxes have been taken from \cite{BP95}, and 
the corresponding theoretical uncertainties have been included in the 
$\chi^2$ covariance matrix as in \cite{Erro}. Asymmetric errors have been 
conservatively symmetrized to the largest one. The $^8$B neutrino energy 
spectrum has been taken from \cite{Bspe}.%
%------------------
\footnote{The $^8$B spectrum shape errors evaluated in \protect\cite{Bspe}
	have not been included in the fits of Figs.~2 and 3, since their 
	effect is much smaller than the uncertainty in the absolute $^8$B 
	neutrino flux. However, such shape errors will be included in the 
	analysis of the energy spectrum performed in Sec.~III and in the 
	Appendix.} 
%------------------
Concerning the neutrino cross sections, we use the updated calculations
of \cite{Bspe} for the chlorine experiment, and of \cite{CrSe} for
the water-Cherenkov experiments. In the MSW calculations, the neutrino 
production regions and the electron density profiles have been taken from 
\cite{BP95}. The Earth regeneration effect is included analytically
as in \cite{LiMo} for each detector latitude. In the vacuum oscillation
calculations, the time average over the year is performed through the 
analytical approach of \cite{Fa97}. In conclusion, Figs.~2 and 3
represent state-of-the-art results in the field of solar neutrino 
oscillations.

%%%%%%%%%%%%%%%%%%%%%%%%%%%%%%%%%%%%%%%%%%%%%%%%%%%%%%%%%%%%%%%%%%%%%%%%%%%%
\section{Solar model independent information}
%%%%%%%%%%%%%%%%%%%%%%%%%%%%%%%%%%%%%%%%%%%%%%%%%%%%%%%%%%%%%%%%%%%%%%%%%%%%

	In this section we analyze the solar model independent information 
about the electron energy spectrum and the time variation of the neutrino 
rate measured by SuperKamiokande. Since these data are already more 
accurate than the corresponding Kamiokande ones, we do not include the 
latter in our analysis.

	We recall that the SuperKamiokande experiment has collected a total
of 4395 solar neutrino events above threshold ($E_e>6.5$ MeV) during 306.3
days of  operation with a fiducial volume of 22.5 kton
\cite{To97,Na97,In97}, corresponding to an observed rate of 0.6377 
events/day/kton. Given the quoted deficit factor of 0.3685 \cite{To97}
(see also Table~I), the expected rate (BP95 model, no oscillation) is 
equal to 1.730 events/day/kton.

%%%%%%%%%%%%%%%%%%%%%%%%%%%%%%%%%%%%%%%%%%%%%%%%%%%%%%%%%%%%%%%%%%%%%%%%%%%%
\subsection{Energy spectrum}

	The SuperKamiokande experiment has measured the energy spectrum of 
recoil electrons from $\nu$-$e$ scattering. The spectrum is given
\cite{To97,Na97,In97} in the form of a 16-bin histogram, each bin 
presenting the ratio between the experimental rate and the theoretical 
rate. Since the theoretical spectrum has not been explicitly presented, we 
rely upon our calculations, using for SuperKamiokande the same inputs 
($^8$B $\nu$ spectrum, cross section,  energy resolution and threshold) 
described in the previous section, and normalizing the results to the total
expected rate of 1.730 events/day/kton. The global information about the 
energy spectrum is reported in Table~II and in Fig.~4.

	Table II shows the relevant spectral data in each of the 16 bins 
(numbered in the first column). As shown in the second column, all bins 
have a 0.5 MeV width, with the exception of the last one, which collects 
all events between 14 and 20 MeV (total energy). The third and fourth 
column report the average total energy $E_i$ (in MeV) and the expected 
event  rate $n_i$ (in events/day/kton/bin) for each bin, in the
absence of oscillations (our calculation). The sum of the entries in the 
fourth column gives the expected rate of 1.730 events/day/kton. 
The fifth column reports the ratio $r_i$ between the experimental and 
theoretical rate in each bin. The upper and lower errors of $r_i$ are
given in the following columns. The sixth and seventh columns report
the quadratic sum of statistical and uncorrelated systematic errors 
$(\sigma'_i)$.  Finally, the correlated systematic errors $(\sigma''_i)$
are given in the last two columns. The  data in columns 6--9 have been 
graphically reduced from the plots shown in \cite{To97,Na97,In97} and thus 
may be subject to slight inaccuracies. The numbers are given with three 
decimal places just to avoid further errors due to truncation and 
round-off. The asymmetry between upper and lower errors is mainly due to 
two slightly different energy calibration procedures \cite{In97}. Shifts 
in the absolute energy scale represent the most important source of 
systematic uncertainties, as was emphasized earlier in \cite{BaLi,BKLi}.

	Figure~4 shows the spectral information in a different way. The 
solid curve represents our calculation of the electron spectrum, using 
the best-fit $^8$B neutrino spectrum given in \cite{Bspe}. The shape of 
the $^8$B neutrino spectrum is subject to some uncertainties, that
have been carefully evaluated in \cite{Bspe}. Therefore, we have also 
used the $\pm3\sigma$ deviated neutrino spectra reported in \cite{Bspe}
to calculate the effect on the electron spectrum (dotted curves). Notice
that in Fig.~4 both the central and the $\pm 3\sigma$ deviated 
theoretical electron spectra are renormalized to the same total area as 
the experimental spectrum (0.6377 events/day/kton). The black circles  
represent the experimental spectrum, with  horizontal error bars spanning 
the bin widths, and vertical error bars representing the $1\sigma$ total 
experimental uncertainties (quadratic sum of statistical and systematic 
errors). It can be seen that the slope of the experimental spectrum is 
slightly more gentle than the theoretical spectrum, the rate observed at 
low (high) energies being somewhat suppressed (enhanced) with respect to 
the expectations. This indication, if confirmed with significantly smaller 
errors, would represent unmistakable evidence for a new, energy-dependent 
process (such as flavor oscillations) affecting solar neutrinos along 
their path to the Earth.

	Since the observed spectrum deviates only slightly from the 
expected one, it is reasonable to try to ``summarize''  the spectral 
information in a single parameter, related in some way to the slope 
deviation. We adopt the approach advocated in Refs.~\cite{BaLi,BKLi,LiMo}, 
where the spectral deformations were expanded in a series of deviations
of the spectral moments (the mean, the variance, etc.) from their standard
values. Given the present experimental uncertainties, it is sufficient
to study the deviations of the first moment, namely, of the average
kinetic energy $\langle T \rangle$ of the electrons, from its standard
(no oscillation) value $\langle T \rangle_0$. This reduction procedure 
has the practical advantage that a single parameter 
$(\Delta \langle T \rangle)$  is used in the fits, instead of 16 bins 
with correlated errors.

	Since the value of $\langle T\rangle$ has not been reported in
\cite{To97,Na97,In97}, we have estimated it from the energy spectrum 
information presented in Table~II and in Fig.~4. The reader is referred 
to the Appendix for a detailed derivation. Our result for the 
fractional deviation $\langle T \rangle/\langle T \rangle_0-1$ is:
%..........................................................................
\begin{equation}
	\frac{\langle T \rangle-\langle T \rangle_0}
	{\langle T \rangle_0}\times 100 
	= 0.99 ^{+2.52}_{-0.96}\ ,
\label{eq:DeltaT}
\end{equation}
%..........................................................................
where the errors represent the total uncertainties at $1\sigma$. This 
result represents the starting point of our analysis of the energy spectra 
deviations due to oscillations. It can be noticed that the observed value 
of $\langle T \rangle$ is about $1\sigma$ higher than the expectations, 
with an upper error larger than the lower one. Therefore, negative 
deviations of $\langle T \rangle$  (i.e., negative ``tilts'' of the 
spectrum slope) will be much more constrained than positive deviations.

%%%%%%%%%%%%%%%%%%%%%%%%%%%%%%%%%%%%%%%%%%%%%%%%%%%%%%%%%%%%%%%%%%%%%%%%%%%%
\subsection{Time variations}

	Possible variations of the neutrino rate during the night would 
represent evidence for the a regeneration of $\nu_e$'s in the Earth
matter, as expected within the MSW scenario (see, e.g., \cite{LiMo}
and references therein). The SuperKamiokande experiment has found no 
such effect within the present uncertainties. In fact, the measured 
asymmetry between  nighttime $(N)$ and daytime $(D)$ neutrino rates is 
\cite{To97,Na97,In97}
%...........................................................................
\begin{equation}
	\frac{N-D}{N+D}=
	0.017\pm0.026({\rm stat.})\pm0.017({\rm syst.})\ .
\label{eq:ND}
\end{equation}
%...........................................................................

	No evidence for variations of the neutrino rate during the time
of the year has been found so far \cite{To97}. Semiannual modulations of 
the signal (in addition to the trivial $1/L^2$ geometrical variations) 
would indicate the presence of neutrino oscillations in vacuum (see, e.g., 
\cite{Fa97,Four} and references therein). A Fourier analysis of the
signal expected in SuperKamiokande  shows that this experiment is 
sensitive only to the first harmonic $f_1$ which, in the absence of 
oscillations, is equal in value to the Earth's orbit eccentricity 
$\varepsilon=0.0167$ (see \cite{Four} for details). The purely statistical 
error $\sigma_f$ of the difference $f_1-\varepsilon$ is given by 
\cite{Four}
%...........................................................................
\begin{equation}
	\sigma_f \simeq \sqrt{\frac{N_S+N_B}{2\,N_S^2}}\ ,
\label{eq:fourier}
\end{equation}
%...........................................................................
where $N_S$ and $N_B$ are the number of signal and background events,
respectively.

	The numbers $N_S$ and $N_B$ depend on the cut applied to the 
fiducial angle in the direction of the sun $(\theta_{\rm sun})$. Using 
Eq.~(\ref{eq:fourier}) and the spectrum of $N_S+N_B$ in terms of 
$\cos\theta_{\rm sun}$ as reported in \cite{To97,Na97,In97}, we derive 
the values of $\sigma_f$ shown in Table~III. The dependence of $\sigma_f$ 
on the cuts applied to $\cos\theta_{\rm sun}$ turns out to be weak, since 
the effect of a lower signal tends to be compensated by a 
a better signal-to-background ratio, and vice versa. On the basis of the
results reported in Table~III, we assume
%...........................................................................
\begin{equation}
	\sigma_f\simeq \pm0.020 {\rm\ (1\ yr)}
\label{eq:typ}
\end{equation}
%...........................................................................
as the typical statistical error of the first Fourier harmonic, after one 
year of operation at SuperKamiokande. The errors corresponding to 2 and 4 
years of operation are simply obtained with rescaling factors of 
$1/\sqrt{2}$ and $1/2$, respectively (assuming the current background 
event rate). Possible systematic errors that might affect
the determination of $f_1$ are not considered here.

%%%%%%%%%%%%%%%%%%%%%%%%%%%%%%%%%%%%%%%%%%%%%%%%%%%%%%%%%%%%%%%%%%%%%%%%%%%%
\subsection{Implications for oscillations in matter}

	Figure~5 illustrates the implications of the electron energy 
spectrum measured by SuperKamiokande [as summarized by the datum of 
Eq.~(\ref{eq:DeltaT})] for the MSW scenario. The upper panel shows curves 
at constant values of the fractional deviation of $\langle T\rangle$ 
expected in the presence of MSW oscillations. Similar curves were 
discussed earlier in \cite{LiMo} (with a different energy threshold) and 
in \cite{BKLi} (without the Earth regeneration effect). Superposed are 
the small and large mixing angle solutions found in Fig.~2 (at 95\% C.L.). 
The central value in Eq.~(\ref{eq:DeltaT}) is intriguingly close to the
values of $\Delta \langle T\rangle/\langle T\rangle_0$ within the small
mixing angle solution, which thus appears slightly favored as compared to
the large mixing one. Unfortunately, the errors in Eq.~(\ref{eq:DeltaT}) 
are still  large, so that only {\em excluded\/} regions can be 
meaningfully derived at present. The lower panel shows the regions 
excluded at 2, 3, and 4 standard deviations by Eq.~(\ref{eq:DeltaT}). For 
instance, in the regions excluded at the  $2\sigma$ level, the theoretical 
value of  $\Delta \langle T\rangle/\langle T\rangle_0$ is either smaller
than $-0.93\%$ or greater than +6.05\%. The excluded regions are similar 
to those shown by the SuperKamiokande Collaboration \cite{To97,Na97,In97}.

	Figure~6 illustrates the implications of the night-day asymmetry 
datum of Eq.~\ref{eq:ND} for the MSW scenario. The upper panel shows 
iso-lines of the night-day asymmetry (percentage) expected in the
presence of oscillations. The lower panel shows the regions excluded
at 2, 3 and 4 standard deviations by the datum of Eq.~\ref{eq:ND}.
The lower part of the large angle solution, where the asymmetry is 
predicted to be $\sim 10\%$, appears to be disfavored by the experimental 
results.

	We do not attempt here a global $\chi^2$-combination of all the 
SuperKamiokande data (total rate, energy spectrum, and night-day 
variations), since these pieces of information are not independent; in 
fact, they are just different ``projections'' of the double differential 
spectrum of events as a function of time and energy. A proper analysis of 
the whole SuperKamiokande data will be possible when such spectrum and its
uncertainties will become available.

	In conclusion, the present data seem to favor the small angle
solution with respect to the large mixing one, because: 1) The global value
of $\chi^2$ is lower at the small mixing solution; 2) The energy spectrum 
shows a slight deviation consistent with small mixing; 3) Part of the 
large mixing angle solution is disfavored by the nonobservation of a 
large night-day asymmetry.

%%%%%%%%%%%%%%%%%%%%%%%%%%%%%%%%%%%%%%%%%%%%%%%%%%%%%%%%%%%%%%%%%%%%%%%%%%%%
\subsection{Implications for oscillations in vacuum}

	Figure~7 illustrates the implications the energy spectrum
measurement [as summarized by the datum in Eq.~(\ref{eq:DeltaT})] for the 
vacuum oscillation scenario. The upper panel shows curves at constant 
values of the fractional deviation of $\langle T\rangle$ expected in the 
presence of vacuum oscillations. Similar curves were discussed earlier in
\cite{BKLi} (with a different energy threshold for SuperKamiokande).
Superposed are the multiple solutions found in the vacuum oscillation fit 
of Fig.~3 (at 95\% C.L.). The  solutions at ``low $\delta m^2$'' 
($\delta m^2 \simeq 6$--$7.5\times 10^{-11}$ eV$^2$) predict positive 
deviations of $\langle T\rangle$, in agreement with the datum of
Eq.~(\ref{eq:DeltaT}). The ``high $\delta m^2$'' solutions predict
negative deviations, and thus are strongly disfavored, as shown in the
lower panel of Fig.~7. Similar results have been discussed by the 
SuperKamiokande Collaboration \cite{To97,Na97,In97}, and are consistent
with earlier Kamiokande spectrum fits \cite{Ha97}.

	Figure~8 illustrates the sensitivity of the SuperKamiokande 
experiment to semiannual modulations of the neutrino rate induced by vacuum
oscillations. The solid curve represents the deviations of the first Fourier
coefficient $f_1$ from its standard value $\varepsilon$ (see \cite{Four}), 
plotted as function of $\delta m^2$ in the most favorable condition (maximal
oscillation amplitude, $\sin^22\theta=1$). The largest vertical error bar 
corresponds to the $\pm1\sigma$ statistical error estimated in 
Eq.~(\ref{eq:typ}) for one year of operation of SuperKamiokande. It can
be seen that the 1~yr error bars are as large as the range spanned by 
variations of $f_1-\varepsilon$, which are thus practically undetectable 
at present.  However, the error bars after 2 and 4 years of data taking
should be reduced enough to allow some sensitivity to the largest 
predicted variations. Of course,  improvements in the energy threshold or 
in the background rate might reduce the uncertainties more rapidly than 
the time sequence shown in Fig.~8.

	By comparing Figs.~7 and 8, it can be noticed that the deviations 
of $\langle T \rangle$ and of $f_1$ have the same sign (both positive or 
negative) at equal values of $\delta m^2$. Such correlation between
energy spectrum deviations and  time variations of the signal can be 
traced to the $L/E$ dependence of the vacuum oscillation probability, as 
recently emphasized in \cite{Mi97}.

	In conclusion, the SuperKamiokande energy spectrum data 
constrain rather strongly the vacuum oscillation solutions, allowing only 
the restricted range $\delta m^2\simeq 6$--$7.5\times 10^{-11}$
eV$^2$ (at large mixing). The present experimental sensitivity
does not allow detection of semiannual modulations of the neutrino rate
due to oscillations. Detection at greater than $1\sigma$ level should
require at least one more year of operation, unless the uncertainties get 
reduced further by improving the energy threshold or the background rate, 
relative to the present levels.

%%%%%%%%%%%%%%%%%%%%%%%%%%%%%%%%%%%%%%%%%%%%%%%%%%%%%%%%%%%%%%%%%%%%%%%%%%%%
\section{Summary and Conclusions}
%%%%%%%%%%%%%%%%%%%%%%%%%%%%%%%%%%%%%%%%%%%%%%%%%%%%%%%%%%%%%%%%%%%%%%%%%%%%

	We have performed a thorough analysis of the solar neutrino
problem in two scenarios: MSW and vacuum oscillations (between two 
neutrino families). We have used the most recent available data, 
including the preliminary results of the SuperKamiokande experiment 
after 306.3 days of operation. Particular attention has been devoted to 
the treatement of the new, solar-model independent data, namely,
the energy spectrum information and the time variations of the signal
in SuperKamiokande. It has been shown that: 1) the small mixing
angle solution is preferred over the large angle one; 2) the vacuum 
oscillation solutions are strongly constrained by the energy spectrum 
measurements; and 3) the detection  of semiannual modulations of the 
$^8$B neutrino flux in SuperKamiokande should require at least one more 
year of operation.

%%%%%%%%%%%%%%%%%%%%%%%%%%%%%%%%%%%%%%%%%%%%%%%%%%%%%%%%%%%%%%%%%%%%%%%%%%%%
\acknowledgments 

	D.M.\ thanks the International School for Advanced Studies
(SISSA, Trieste, Italy), and in particular A.\ Masiero and S.\ T.\ Petcov, 
for kind hospitality during the early stages of this work. E.L.\ 
thanks K.\ Inoue and A.\ Yu.\ Smirnov for helpful conversations during 
the TAUP'97 Workshop.

%%%%%%%%%%%%%%%%%%%%%%%%%%%%%%%%%%%%%%%%%%%%%%%%%%%%%%%%%%%%%%%%%%%%%%%%%%%%
\appendix\section*{Evaluation of $\langle T\rangle$ 
	from the SuperKamiokande spectrum}
%%%%%%%%%%%%%%%%%%%%%%%%%%%%%%%%%%%%%%%%%%%%%%%%%%%%%%%%%%%%%%%%%%%%%%%%%%%%

	In this Appendix we give the detailed derivation of the result 
reported in Eq.~(\ref{eq:DeltaT}). We make use of the preliminary 
SuperKamiokande spectrum data after 306.3 days of operation 
\cite{To97,Na97,In97}, as reported in Table~II and shown in Fig.~4.

	The average value of the total electron energy $E_e$ from the 
unbinned theoretical spectrum $n(E_e)$ (solid curve in Fig.~4) is given by
%...........................................................................
\begin{equation}
	\langle E_e \rangle_{\rm theo}({\rm unbinned})=
	\frac{\displaystyle
	\int_{\rm 6.5\rm\ MeV}^{20\rm  MeV}dE_e\;n(E_e)\,E_e}
	{\displaystyle
	\int_{\rm 6.5\rm\ MeV}^{20\rm  MeV}dE_e\;n(E_e)}=8.536\rm\ MeV\ .
\label{eq:Eunbinned}
\end{equation}
%...........................................................................
Using instead the binned theoretical spectrum $n_i$ from Table~II one 
obtains:
%...........................................................................
\begin{equation}
\label{eq:Ebinned}
\langle E_e \rangle_{\rm theo}({\rm binned})=
\frac{\sum_{i=1}^{16}n_i\,E_i}
{\sum_{i=1}^{16}n_i}=8.525\rm\ MeV\ .
\end{equation}
%...........................................................................
The difference between the values in Eqs.~(\ref{eq:Eunbinned}) and 
(\ref{eq:Ebinned}) is only $\sim 0.1\%$, indicating that the effect of 
binning is not important in the evaluation of the average energy. We will 
anyway attach to  $\langle E_e \rangle_{\rm theo}$ a $\pm 0.010$ MeV 
``binning error'' that, as we shall see, is an order of magnitude smaller 
than the present experimental uncertainties. Another small error is 
induced by the $^8$B neutrino spectrum uncertainty (as shown by the 
dotted curves in Fig.~4), that we evaluate as 
$\Delta\langle E_e\rangle_{\rm theo}=\pm0.016$ MeV at the $1\sigma$ level.

	The average value of $E_e$ derived from the (binned) experimental 
spectrum is given by
%...........................................................................
\begin{equation}
	\langle E_e\rangle_{\rm exp} = 
	\frac{\sum_{i=1}^{16}n_i\,r_i\,E_i}
	{\sum_{i=1}^{16}n_i\,r_i}=8.604\rm\ MeV\ ,
\label{eq:Eexp}
\end{equation}
%...........................................................................
where $r_i$ and $E_i$ are given in Table~II. In general, the squared error 
$\sigma^2_{\langle E\rangle}$ associated to $\langle E_e\rangle_{\rm exp}$ 
reads
%...........................................................................
\begin{equation}
 	\sigma^2_{\langle E\rangle}=
 	\sum_{i=1}^{16}\sum_{j=1}^{16}
 	\frac{\partial \langle E_e\rangle_{\rm exp}}{\partial r_i}\,
	\frac{\partial \langle E_e\rangle_{\rm exp}}{\partial r_j}\,
	\rho_{ij}\,\sigma_i\,\sigma_j\ ,
\label{eq:sigma}
\end{equation}
%...........................................................................
where $\sigma_i$ represents the error of $r_i$ and $\rho_{ij}$ is the
correlation matrix. The partial derivatives are easily obtained from 
Eq.~(\ref{eq:Eexp}):
%...........................................................................
\begin{equation}
	\frac{\partial \langle E_e\rangle_{\rm exp}}{\partial r_i}
	= n_i\,\frac{E_i-\langle E\rangle_{\rm exp}}
	{\sum_{j=1}^{16}n_j\,r_j}\ .
\label{eq:deriv}
\end{equation}
%...........................................................................

	The correlation coefficients for the statistical and uncorrelated
systematic errors are given by $\rho_{ij}=\delta_{ij}$, so that 
from Eq.~(\ref{eq:sigma}) one obtains
%...........................................................................
\begin{equation}
	\sigma_{\langle E\rangle}({\rm stat.+uncorr.~syst.})=\pm
	\left[\sum_{i=1}^{16}
	\left(\frac{\partial \langle E_e\rangle_{\rm exp}}
	{\partial r_i}\,\sigma'_i
	\right)^2\right]^{\frac{1}{2}}=\left\{
	\begin{array}{c}+0.108\\-0.055\end{array}\right.
	{\rm\ (MeV)}\ ,
\label{eq:sigmaunc}
\end{equation}
%...........................................................................
where the $\sigma_i'$'s are taken from Table~II, upper and lower errors 
being propagated separately. The remaining systematic errors are fully
correlated \cite{In97}, i.e., $\rho_{ij}=1$, so that
%...........................................................................
\begin{equation}
	\sigma_{\langle E\rangle}({\rm corr.~syst.})=\pm
	\sum_{i=1}^{16}
	\frac{\partial \langle E_e\rangle_{\rm exp}}
	{\partial r_i}\,\sigma''_i
	=\left\{
	\begin{array}{c}+0.170\\-0.050\end{array}\right.
	{\rm\ (MeV)}\ ,
\label{eq:sigmacor}
\end{equation}
%...........................................................................
where the $\sigma''_i$'s are given in the last two columns of Table~II.

	Taking the difference between Eqs.~(\ref{eq:Eexp}) and 
(\ref{eq:Ebinned}) and including all the uncertainties, one obtains
%...........................................................................
\begin{equation}
	\langle E\rangle_{\rm exp} - \langle E\rangle_{\rm theo} =
	0.079\  
	{}^{+0.108}_{-0.055}\ 
	{}^{+0.170}_{-0.050}\ 
	\pm 0.010\ 
	\pm 0.016\ ({\rm MeV})\ ,
\label{eq:DeltaE}
\end{equation}
%...........................................................................
where the first error is due to statistics and to uncorrelated systematics,
the second to correlated systematics, the third to the electron spectrum
binning, and the fourth to the $^8$B neutrino spectrum uncertainties. In 
conclusion, passing from total to kinetic energies, it follows that 
%...........................................................................
\begin{equation}
	\frac{\langle T \rangle_{\rm exp}-\langle T \rangle_{\rm theo}}
	{\langle T \rangle_{\rm theo}} = 0.99 ^{+2.52}_{-0.96}\ \%\ ,
\label{eq:DeltaTapp}
\end{equation}
%...........................................................................
having added in quadrature the $1\sigma$ independent errors 
in Eq.~(\ref{eq:DeltaE}). This is the final result anticipated in 
Eq.~(\ref{eq:DeltaT}).

%%%%%%%%%%%%%%%%%%%%%%%%%%%%%%%%%%%%%%%%%%%%%%%%%%%%%%%%%%%%%%%%%%%%%%%%%%
%		               T A B L E S
%%%%%%%%%%%%%%%%%%%%%%%%%%%%%%%%%%%%%%%%%%%%%%%%%%%%%%%%%%%%%%%%%%%%%%%%%%

%%%%%%%%%%%%%%%%%%%%%%%%%%%%%%%%%%%%%%%%%%%%%%%%%%%%%%%%%%%%%%%%%%%%%%%%%%
%%%%%%%%%%%%%%%%%%%%%%%%  TABLE I  %%%%%%%%%%%%%%%%%%%%%%%%%%%%%%%%%%%%%%%
\begin{table}
\caption{Neutrino event rates measured by solar neutrino experiments,
	and corresponding predictions from the BP95 standard solar model.
	The quoted errors are at $1\sigma$.}
\begin{tabular}{ccccc}
%=========================================================================
Experiment 	&Ref.& Data & Theory \protect\cite{BP95}& Units \\
\tableline
%-------------------------------------------------------------------------
Homestake	& \protect\cite{La97}& $ 2.54 \pm 0.16 \pm 0.14$         &
	$9.3^{+1.2}_{-1.4}$    & SNU					\\
Kamiokande 	& \protect\cite{Fu96}& $ 2.80 \pm 0.19 \pm 0.23$         &
	$6.62^{+0.93}_{-1.12}$ & $10^6$ cm$^{-2}~$s$^{-1}$		\\
SAGE		& \protect\cite{SAGE}& $   73 \pm  8.5   ^{+5.2} _{-6.9}$&
	$137^{+8}_{-7}$        & SNU					\\
GALLEX		& \protect\cite{Cr97}& $ 76.2 \pm  6.5 \pm    5$         &
	$137^{+8}_{-7}$	       & SNU					\\
SuperKamiokande	& \protect\cite{To97}& $ 2.44 \pm 0.06  ^{+0.25}_{-0.09}$&
 	$6.62^{+0.93}_{-1.12}$ & $10^6$ cm$^{-2}~$s$^{-1}$	
%=========================================================================
\end{tabular}
\end{table}

%%%%%%%%%%%%%%%%%%%%%%%%%%%%%%%%%%%%%%%%%%%%%%%%%%%%%%%%%%%%%%%%%%%%%%%%%%%
%%%%%%%%%%%%%%%%%%%%%%%  TABLE II  %%%%%%%%%%%%%%%%%%%%%%%%%%%%%%%%%%%%%%%%
\begin{table}
\caption{The electron energy spectrum at SuperKamiokande. First three 
	columns: sequential number, energy range, and average energy for 
	each bin. Fourth column: expected number of events $n_i$ per bin 
	per day per kiloton  without oscillations (our calculation). Fifth 
	column: ratio of measured to expected rates in each bin. Sixth and 
	seventh columns: Upper and lower $1\sigma$ errors of $r_i$ from 
	statistical and uncorrelated systematic uncertainties. Eighth and 
	ninth columns: Upper and lower $1\sigma$ errors of $r_i$ from 
	correlated systematic uncertainties. The numbers in columns 6--9 
	have been graphically reduced from the plots shown in 
	\protect\cite{To97,Na97,In97}.} 
\begin{tabular}{ccccccccc}
%=========================================================================
$i$						  &   
Range \tablenotemark[1]    			  & 
$E_i$ \tablenotemark[1] 			  & 
$n_i$ \tablenotemark[2]				  & 
$r_i$               				  & 
\multicolumn{2}{c}{$\pm\sigma'_i$ (stat.+unc.)}   & 
\multicolumn{2}{c}{$\pm\sigma''_i$ (corr.\ syst.)}\\
\tableline
%-------------------------------------------------------------------------
 1 & $[6.5,\,7]$ &  6.747& 0.309& 0.346& +0.046&$-$0.032&+0.010&$-$0.004\\
 2 & $[7,\,7.5]$ &  7.244& 0.271& 0.341& +0.045&$-$0.032&+0.013&$-$0.006\\
 3 & $[7.5,\,8]$ &  7.742& 0.232& 0.350& +0.043&$-$0.036&+0.016&$-$0.008\\
 4 & $[8,\,8.5]$ &  8.239& 0.198& 0.400& +0.062&$-$0.037&+0.025&$-$0.010\\
 5 & $[8.5,\,9]$ &  8.742& 0.165& 0.327& +0.064&$-$0.039&+0.026&$-$0.012\\
 6 & $[9,\,9.5]$ &  9.240& 0.135& 0.373& +0.075&$-$0.046&+0.040&$-$0.014\\
 7 & $[9.5,\,10]$&  9.736& 0.109& 0.382& +0.092&$-$0.049&+0.050&$-$0.016\\
 8 &$[10,\,10.5]$& 10.239& 0.087& 0.373& +0.101&$-$0.053&+0.058&$-$0.017\\
 9 &$[10.5,\,11]$& 10.737& 0.066& 0.356& +0.115&$-$0.059&+0.069&$-$0.023\\
10 &$[11,\,11.5]$& 11.234& 0.050& 0.372& +0.138&$-$0.062&+0.089&$-$0.026\\
11 &$[11.5,\,12]$& 11.736& 0.036& 0.392& +0.170&$-$0.072&+0.110&$-$0.036\\
12 &$[12,\,12.5]$& 12.234& 0.026& 0.402& +0.209&$-$0.088&+0.137&$-$0.042\\
13 &$[12.5,\,13]$& 12.730& 0.017& 0.359& +0.229&$-$0.098&+0.141&$-$0.046\\
14 &$[13,\,13.5]$& 13.232& 0.012& 0.422& +0.238&$-$0.123&+0.192&$-$0.063\\
15 &$[13.5,\,14]$& 13.730& 0.007& 0.627& +0.506&$-$0.183&+0.337&$-$0.101\\
16 & $[14,\,20]$ & 14.817& 0.010& 0.493& +0.510&$-$0.170&+0.378&$-$0.114
%=========================================================================
\end{tabular}
\tablenotetext[1]{Units: MeV (total electron energy).}
\tablenotetext[2]{Units: events/day/kton/bin. }
\end{table}

%%%%%%%%%%%%%%%%%%%%%%%%%%%%%%%%%%%%%%%%%%%%%%%%%%%%%%%%%%%%%%%%%%%%%%%%%%%
%%%%%%%%%%%%%%%%%%%%%%%%  TABLE III  %%%%%%%%%%%%%%%%%%%%%%%%%%%%%%%%%%%%%%
\begin{table}
\caption{Number of signal events $N_S$ and signal-to-background ratio 
	$N_S/N_B$ for various $\cos\theta_{\rm sun}$ thresholds (as 
	derived from the  $\cos\theta_{\rm sun}$ event spectrum presented 
	in \protect\cite{To97,Na97,In97}), together with the corresponding 
	uncertainty $\sigma_f$ of the deviation of the first Fourier 
	coefficient $f_1$ from its standard value $\varepsilon$ 
	[see Eq.~(\ref{eq:fourier})].}
\begin{tabular}{cccc}
%==========================================================================
$\cos\theta_{\rm sun}$ & $N_S$ & $N_S/N_B$ & $\sigma_f$ \\
\tableline
%--------------------------------------------------------------------------
$>0.5$ & $\sim 4100$ & $\sim 0.30$ & 0.023 \\
$>0.6$ & $\sim 3870$ & $\sim 0.35$ & 0.022 \\
$>0.7$ & $\sim 3620$ & $\sim 0.38$ & 0.021 \\
$>0.8$ & $\sim 3270$ & $\sim 0.59$ & 0.020 \\
$>0.9$ & $\sim 2530$ & $\sim 0.92$ & 0.020 \\
%==========================================================================
\end{tabular}
\end{table}

%%%%%%%%%%%%%%%%%%%%%%%%%%%%%%%%%%%%%%%%%%%%%%%%%%%%%%%%%%%%%%%%%%%%%%%%%%%
% 			R E F E R E N C E S 
%%%%%%%%%%%%%%%%%%%%%%%%%%%%%%%%%%%%%%%%%%%%%%%%%%%%%%%%%%%%%%%%%%%%%%%%%%%

%\end{document}

%%%%%%%%%%%%%%%%%%%%%%%%%%%%%%%%%%%%%%%%%%%%%%%%%%%%%%%%%%%%%%%%%%%%%%%%%%%
%		F I G U R E S 
%%%%%%%%%%%%%%%%%%%%%%%%%%%%%%%%%%%%%%%%%%%%%%%%%%%%%%%%%%%%%%%%%%%%%%%%%%%

%..........................................................................
\begin{figure}
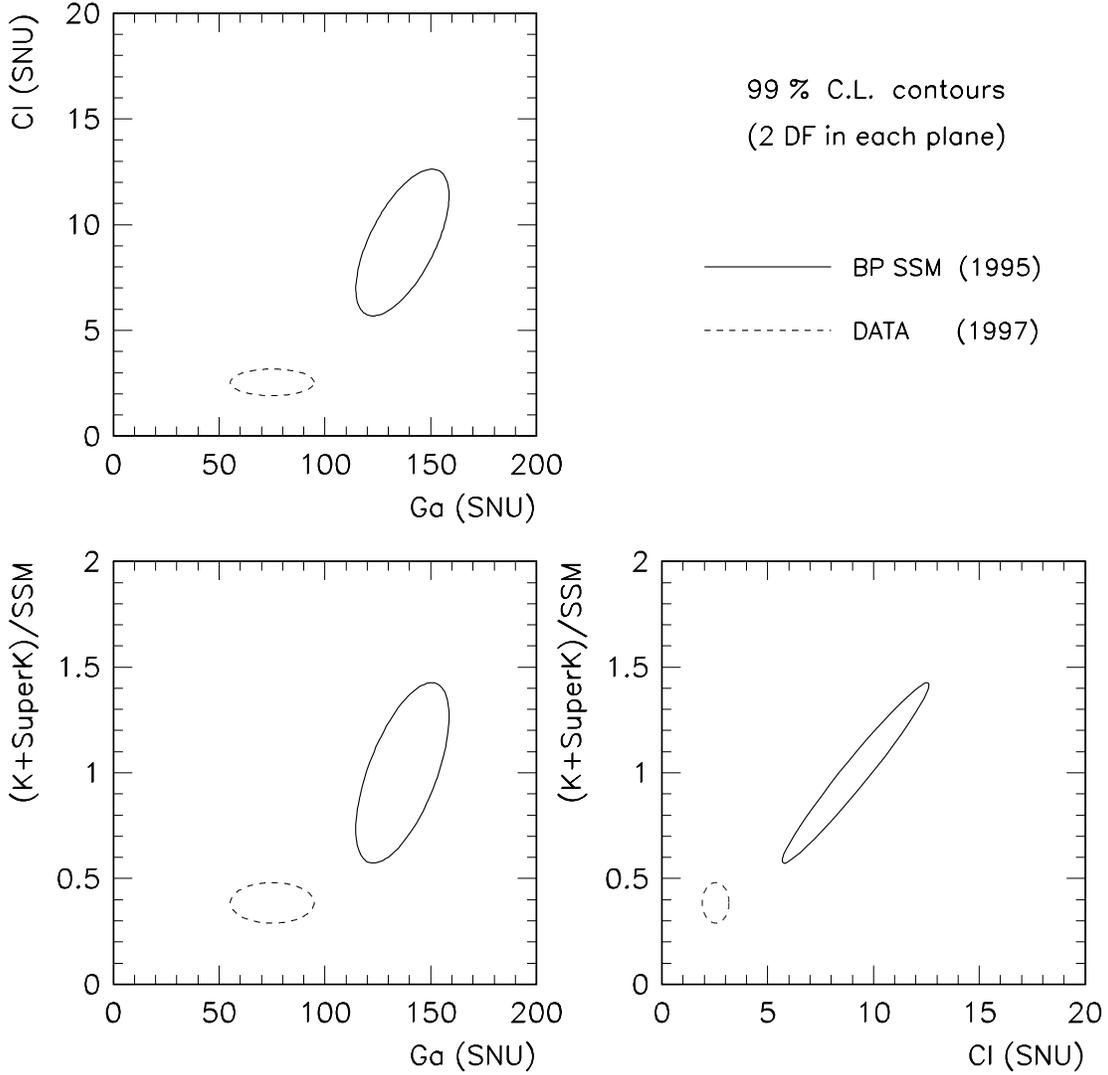

\caption{The current state of the solar neutrino deficit. The ellipses
	represent the regions allowed at 99\% C.L.\ by the present solar 
	neutrino data (dashed lines) and by the standard solar model (SSM) 
	of  Bahcall-Pinsonneault \protect\cite{BP95}. The coordinates are
	the chlorine (Cl), gallium (Ga), and water-Cherenkov [normalized
	(K+SuperK)/SSM] neutrino rates.}
\label{fig:1}
\end{figure}
%..........................................................................
\begin{figure}
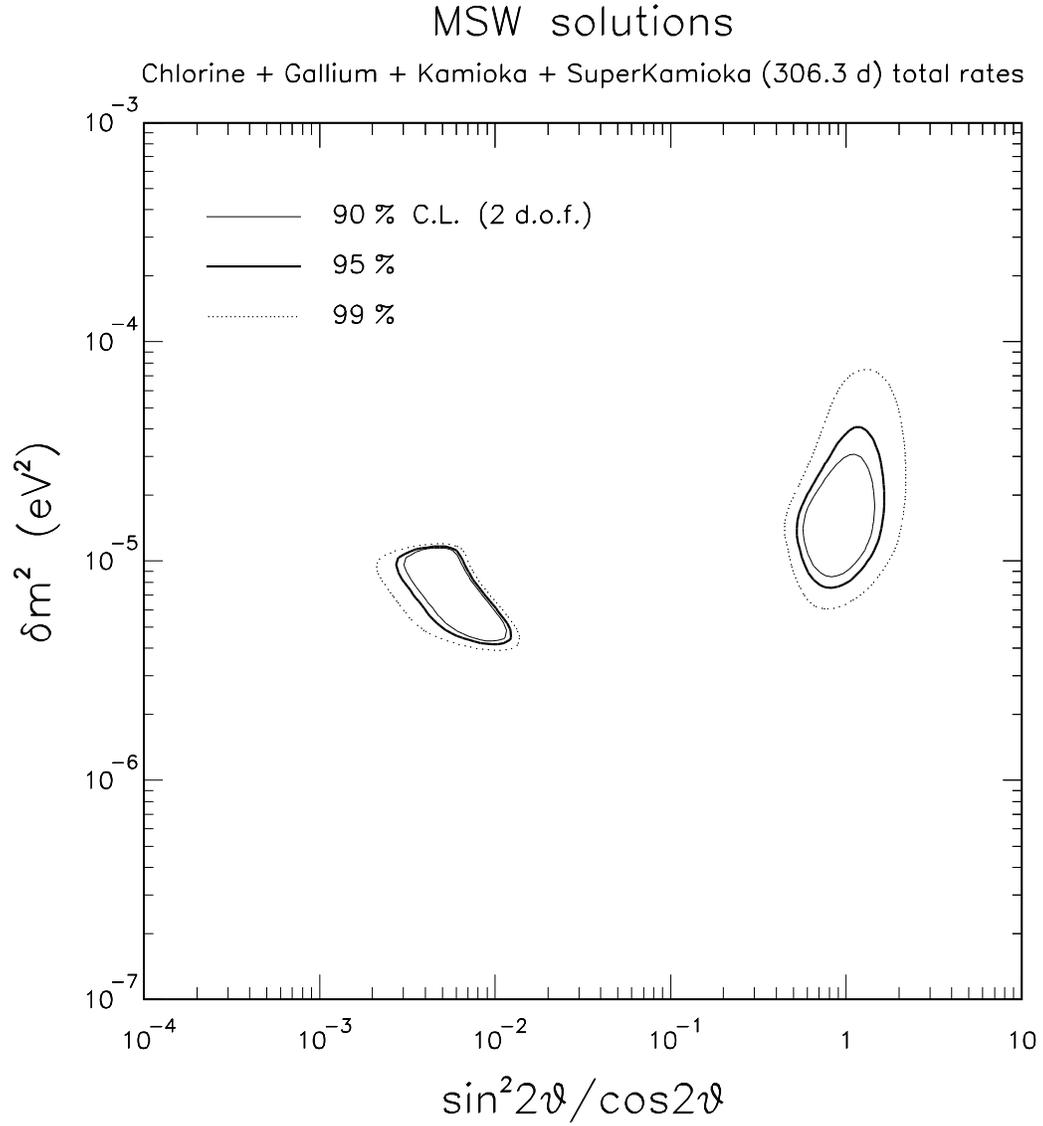

\caption{The MSW small and large mixing angle solutions to the solar
	neutrino problem, as obtained by a fit to the total rates only 
	(energy spectrum and night-day data not included).}
\label{fig:2}
\end{figure}
%..........................................................................
\begin{figure}
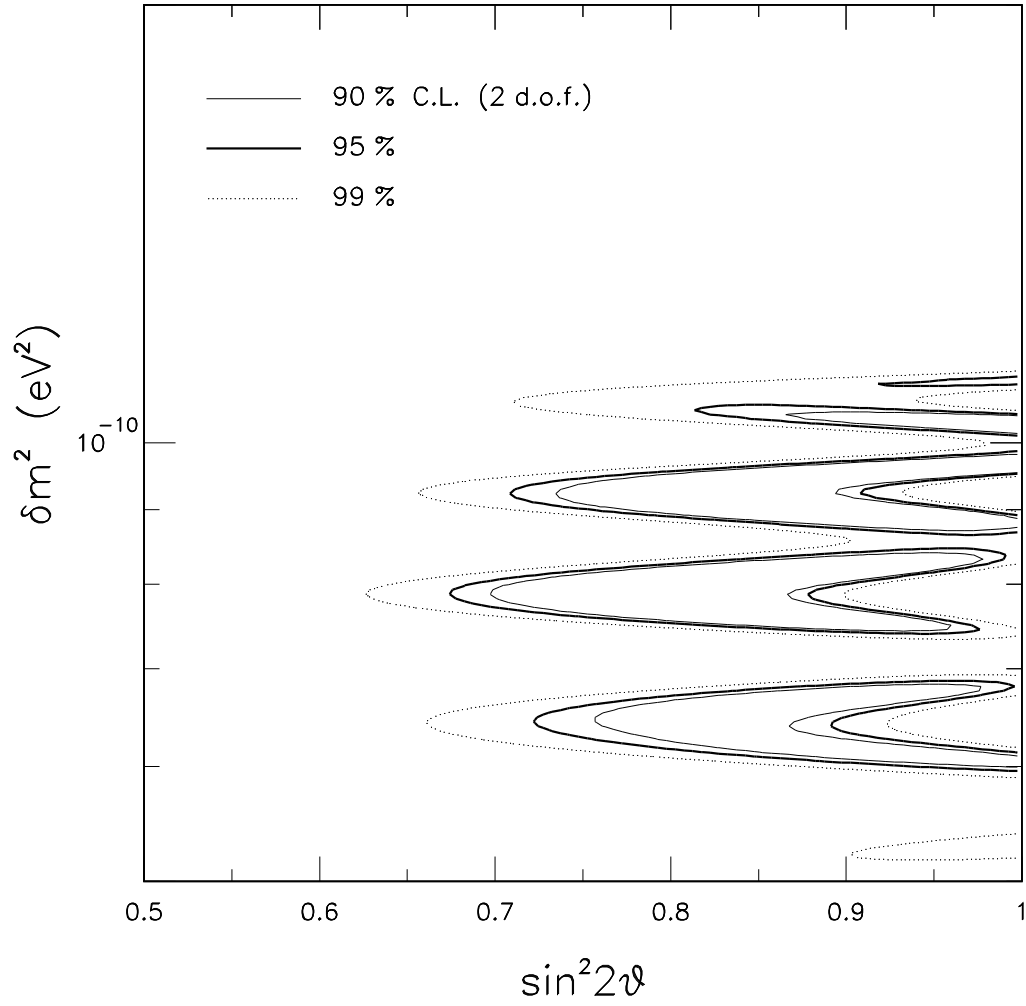

\caption{The vacuum oscillation solutions to the solar
	neutrino problem, as obtained by a fit to the total rates only 
	(energy spectrum and night-day data not included). The range
	of the (logarithmic) vertical scale is (0.5--2)$\times 10^{-10}$ 
	eV$^2$. }
\label{fig:3}
\end{figure}
%..........................................................................
\begin{figure}
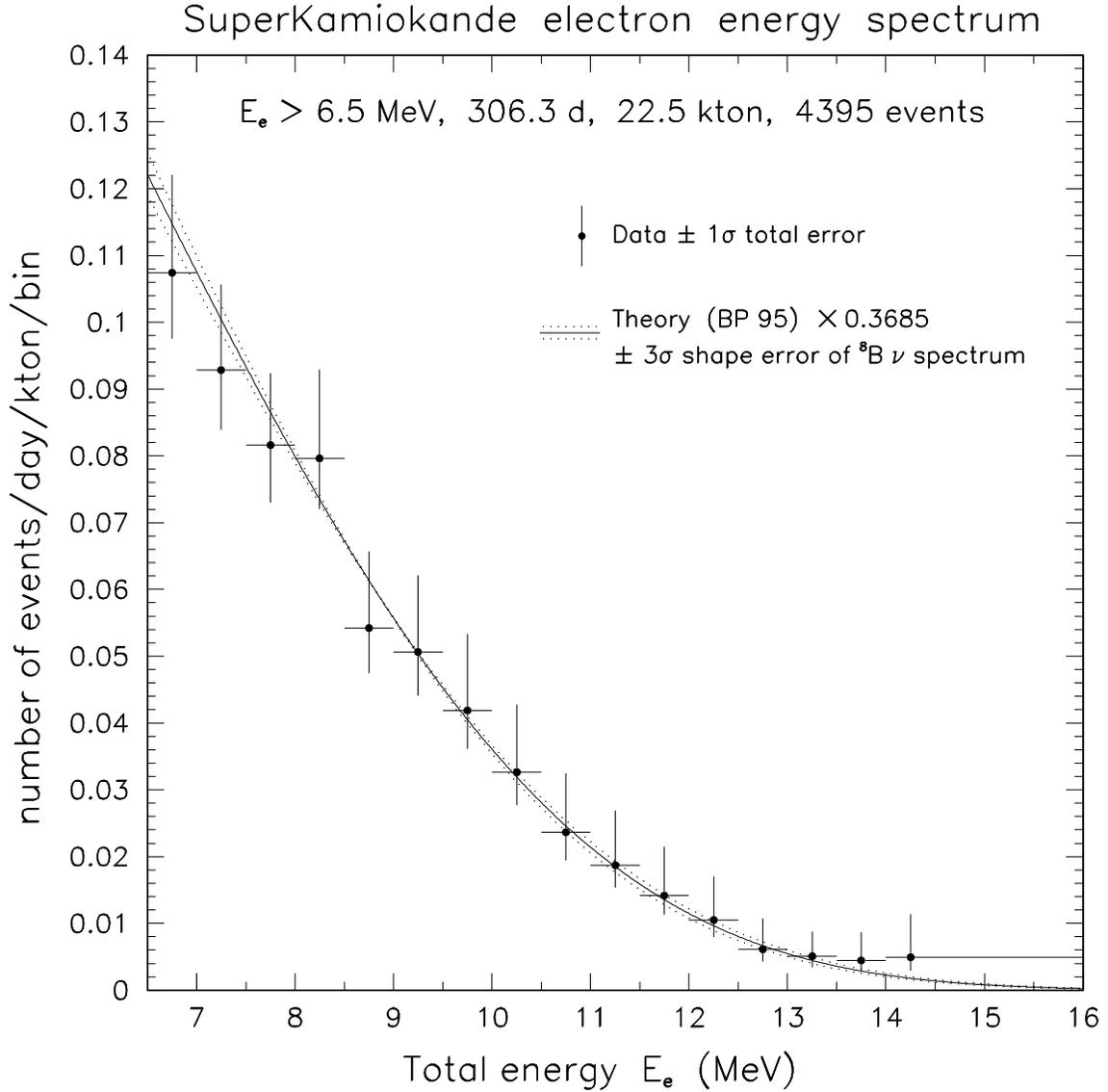

\caption{The electron recoil energy spectrum at SuperKamiokande. The dots 
	represent the measured rates in each energy bin, with vertical 
	$1\sigma$ error bars given by the quadratic sum of statistical and 
	systematic errors. The horizontal bars span the bin widths. Also 
	shown is the standard spectrum expected in the absence of 
	oscillations (solid line), together with the spectra obtained by 
	assuming $\pm 3\sigma$ deviations of the standard $^8$B neutrino 
	spectrum \protect\cite{Bspe}. The data are taken from 
	\protect\cite{To97,Na97,In97}. The theoretical spectra refer
	to our calculations, renormalized to give the same area as the
	experimental histogram.}
\label{fig:4}
\end{figure}
%..........................................................................
\begin{figure}
\caption{Constraints on the MSW solutions as derived by the measured value
	of the average electron kinetic energy $\langle T\rangle$. Upper 
	panel: Theoretical expectations for the fractional shift of 
	$\langle T\rangle$ from its standard (no oscillation) value 
	$\langle T \rangle_0$. Lower panel: regions excluded at 2, 3, 
	and 4 standard deviations by the SuperKamiokande determination 
	of $\langle T\rangle$.}
\label{fig:5}
\end{figure}
%..........................................................................
\begin{figure}
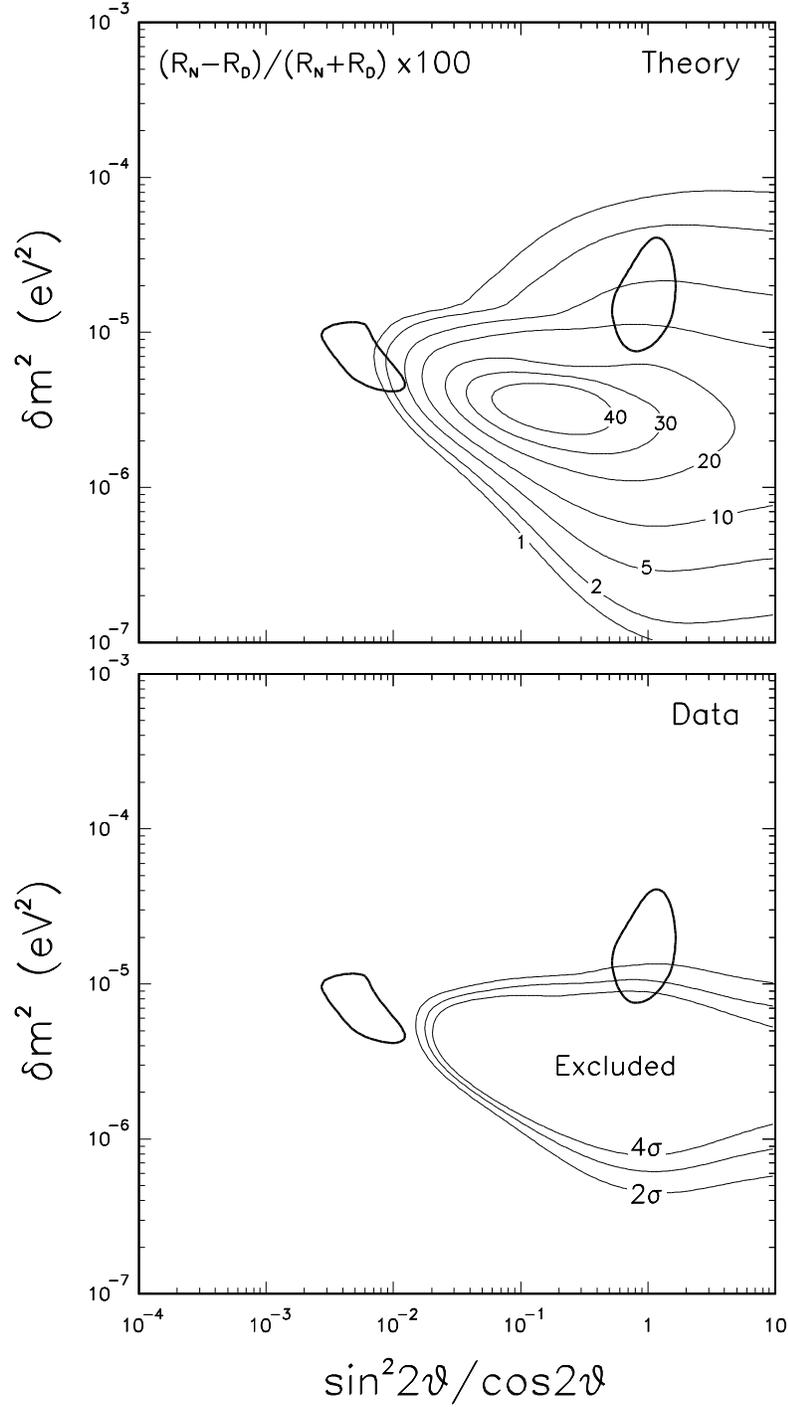

\caption{Constraints on the MSW solutions as derived by the measured
	value of the asymmetry of nighttime ($R_{\rm N}$) and daytime 
	($R_{\rm D}$) event rates. Upper panel: theoretical expectations 
	for $(R_{\rm N}-R_{\rm D})/(R_{\rm N}+R_{\rm D})$. Lower panel: 
	Regions excluded at 2, 3, and 4 standard deviations by the 
	SuperKamiokande data.}
\label{fig:6}
\end{figure}
%..........................................................................
\begin{figure}
\caption{Constraints on the vacuum oscillation solutions as derived by the 
	measured value of the average electron kinetic energy 
	$\langle T\rangle$. Upper panel: Theoretical expectations for the 
	fractional shift of $\langle T\rangle$ from its standard 
	(no oscillation) value $\langle T \rangle_0$. Lower panel: Regions 
	excluded at 2, 3, and 4 standard deviations by the SuperKamiokande 
	determination of $\langle T\rangle$.}
\label{fig:7}
\end{figure}
%..........................................................................
\begin{figure}
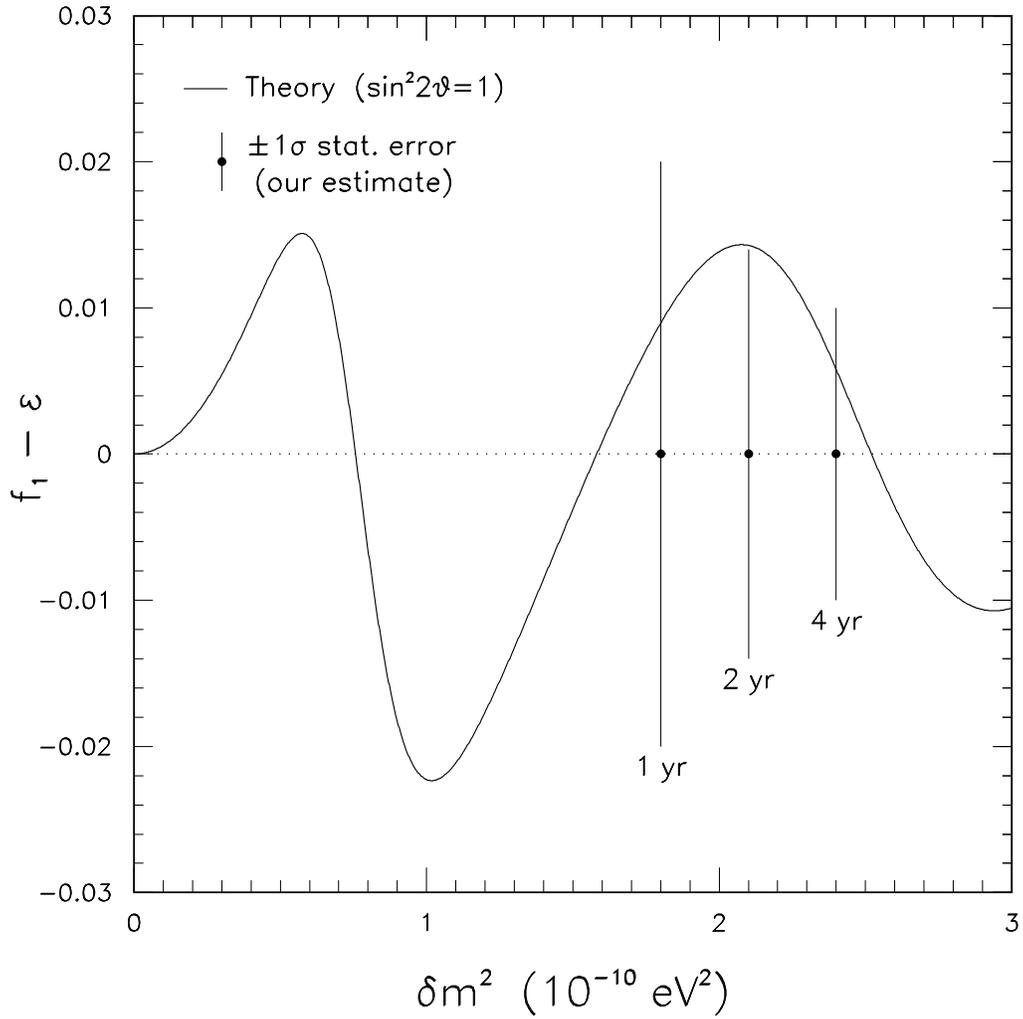

\caption{Fourier analysis of time variations induced by vacuum 
	oscillations. Solid curve: expected deviations of the first 
	Fourier coefficient $f_1$ from its standard value (equal to Earth 
	orbit eccentricity $\varepsilon$) for $\sin^22\theta=1$. Error 
	bars: Our estimate of the statistical errors of SuperKamiokande 
	(including background) after 1, 2, and 4 years of data taking. 
	See the text and \protect\cite{Four} for details.}
\label{fig:8}
\end{figure}
%..........................................................................

%\end{document}

%%%%%%%%%%%%%%%%%%%%%%%%%%%%%%%%%%%%%%%%%%%%%%%%%%%%%%%%%%%%%%%%%%%%%%%%%%%
%%%%%%%%%%%%%%%%%%%%%%%%%%%%%%%%%%%%%%%%%%%%%%%%%%%%%%%%%%%%%%%%%%%%%%%%%%%
%%%%%%% 
%%%%%%%            THE FOLLOWING FOR AUTHOR USE ONLY.
%%%%%%%
%%%%%%%            INCLUSION OF FIGURES WITH EPSFIG.STY.
%%%%%%%
%%%%%%%%%%%%%%%%%%%%%%%%%%%%%%%%%%%%%%%%%%%%%%%%%%%%%%%%%%%%%%%%%%%%%%%%%%
%%%%%%%          P O S T S C R I P T       F I G U R E S 
%%%%%%%
%%%%%%%   memo:  1) add epsfig in the \documentstyle
%%%%%%%          2) and move this part before \end{document} 
%%%%%%%		 3) remove previous figure captions
%%%%%%%          4) include the following \newcommand:
%%-------------------------------------------------------------------------
\newcommand{\InsertFigure}[2]{\newpage\begin{center}\mbox{%
\epsfig{bbllx=1.4truecm,bblly=1.3truecm,bburx=19.5truecm,bbury=26.5truecm,%
height=21.truecm,figure=#1}}\end{center}\vspace*{-1.85truecm}%
\parbox[t]{\hsize}{\small\baselineskip=0.5truecm\hskip0.5truecm #2}}
%--------------------------------------------------------------------------
%%%%%%%%%%%%%%%%%%%%%%%%%%%%%%%%%%%%%%%%%%%%%%%%%%%%%%%%%%%%%%%%%%%%%%%%%%%

%..........................................................................
\InsertFigure{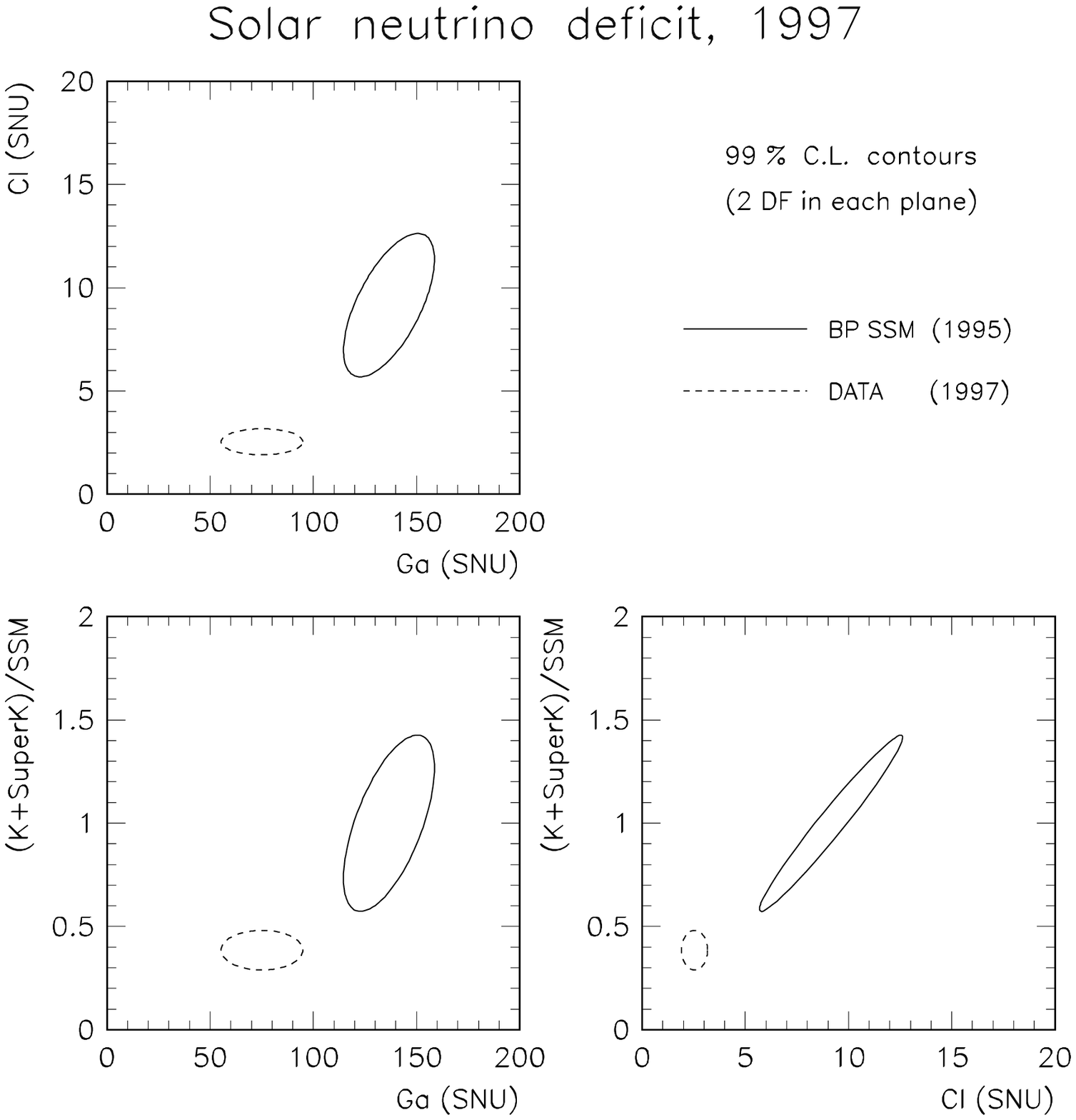}%
{FIG.~1. The current state of the solar neutrino deficit. The ellipses
	represent the regions allowed at 99\% C.L.\ by the present solar 
	neutrino data (dashed lines) and by the standard solar model (SSM) 
	of Bahcall-Pinsonneault \protect\cite{BP95}. The coordinates are
	the chlorine (Cl), gallium (Ga), and water-Cherenkov [normalized
	(K+SuperK)/SSM] neutrino rates.}
%..........................................................................
\InsertFigure{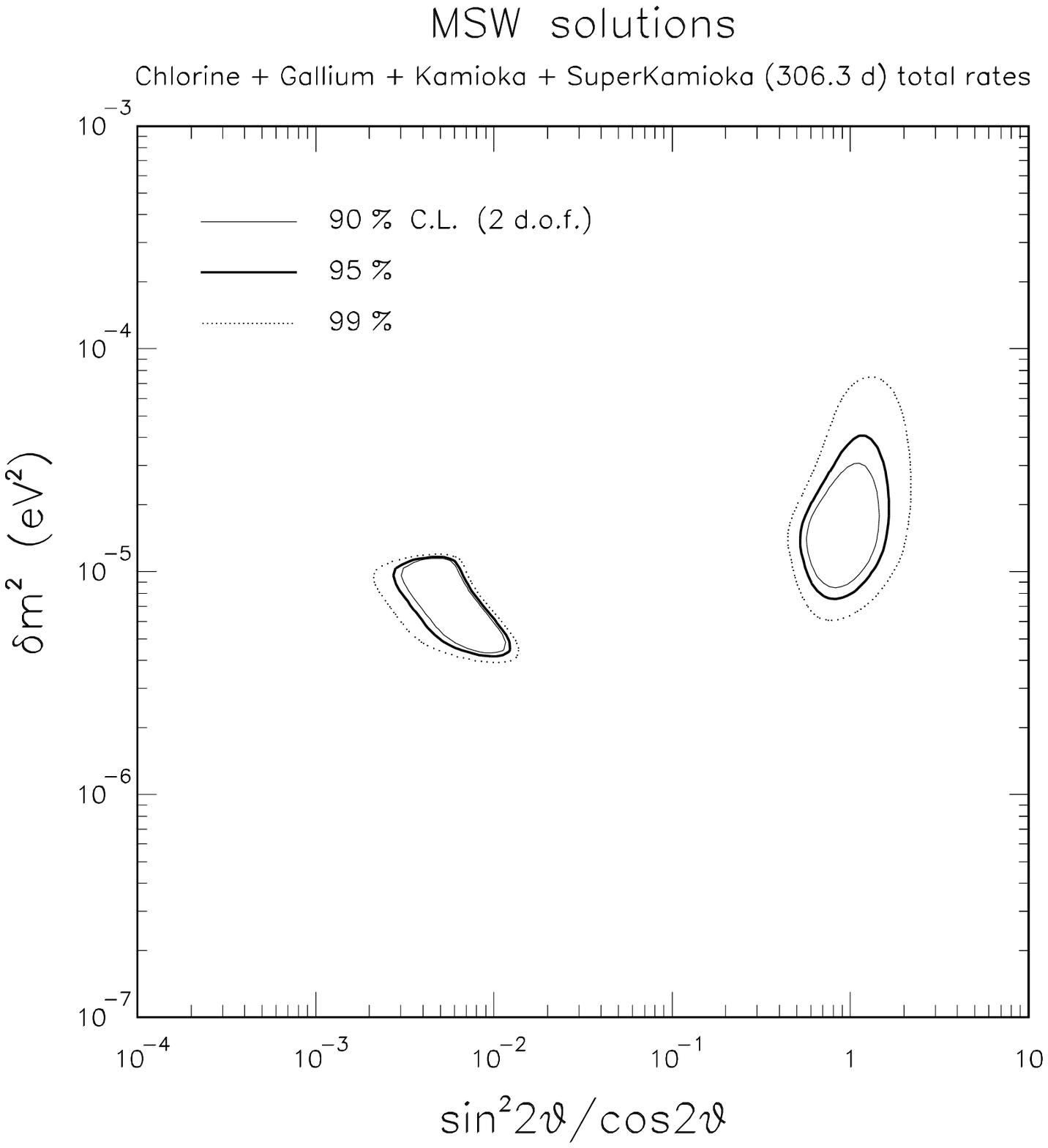}%
{FIG.~2.  The MSW small and large mixing angle solutions to the solar
	neutrino problem, as obtained by a fit to the total rates only 
	(energy spectrum and night-day data not included).}
%..........................................................................
\InsertFigure{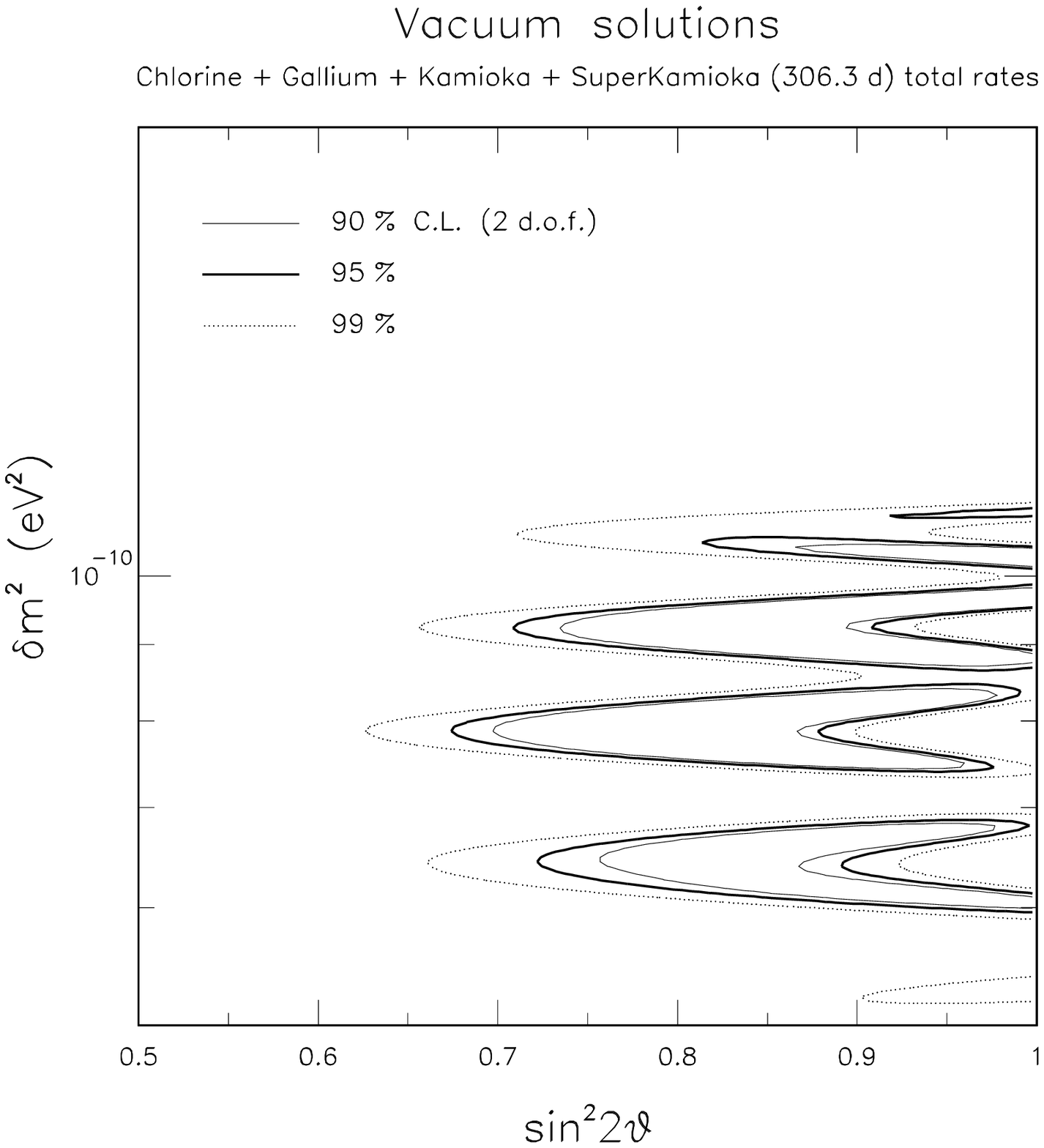}%
{FIG.~3. The vacuum oscillation solutions to the solar
	neutrino problem, as obtained by a fit to the total rates only 
	(energy spectrum and night-day data not included). The range
	of the (logarithmic) vertical scale is (0.5--2)$\times 10^{-10}$ 
	eV$^2$. }
%..........................................................................
\InsertFigure{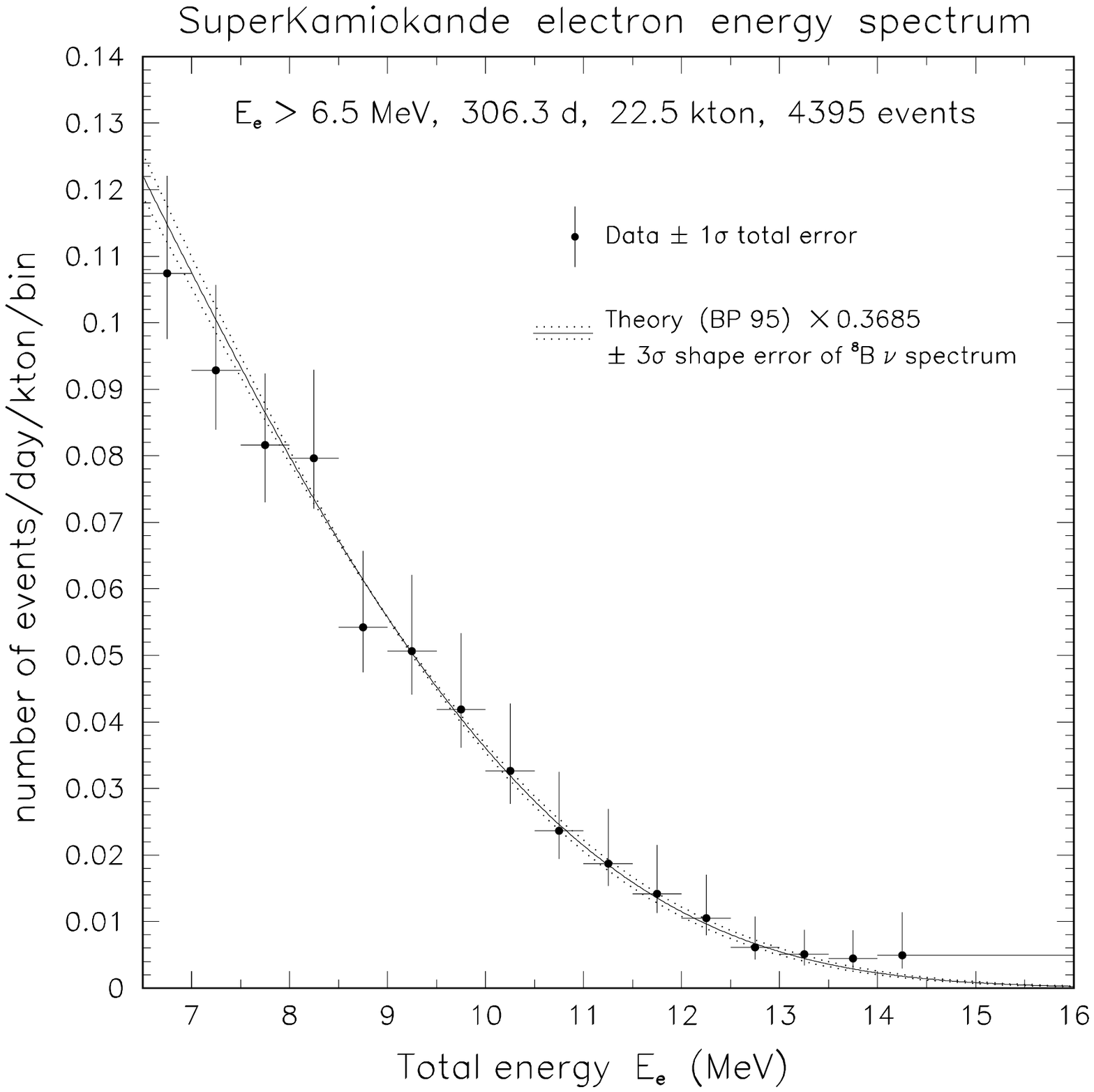}%
{FIG.~4.  The electron recoil energy spectrum at SuperKamiokande. The dots 
	represent the measured rates in each energy bin, with vertical 
	$1\sigma$ error bars given by the quadratic sum of statistical and 
	systematic errors. The horizontal bars span the bin widths. Also 
	shown  is the standard spectrum expected in the absence of 
	oscillations  (solid line), together with the spectra obtained by 
	assuming  $\pm 3\sigma$ deviations of the standard $^8$B neutrino 
	spectrum \protect\cite{Bspe}. The data are taken from 
	\protect\cite{To97,Na97,In97}. The theoretical spectra refer
	to our calculations, renormalized to give the same area as the
	experimental histogram.}
%...........................................................................
\InsertFigure{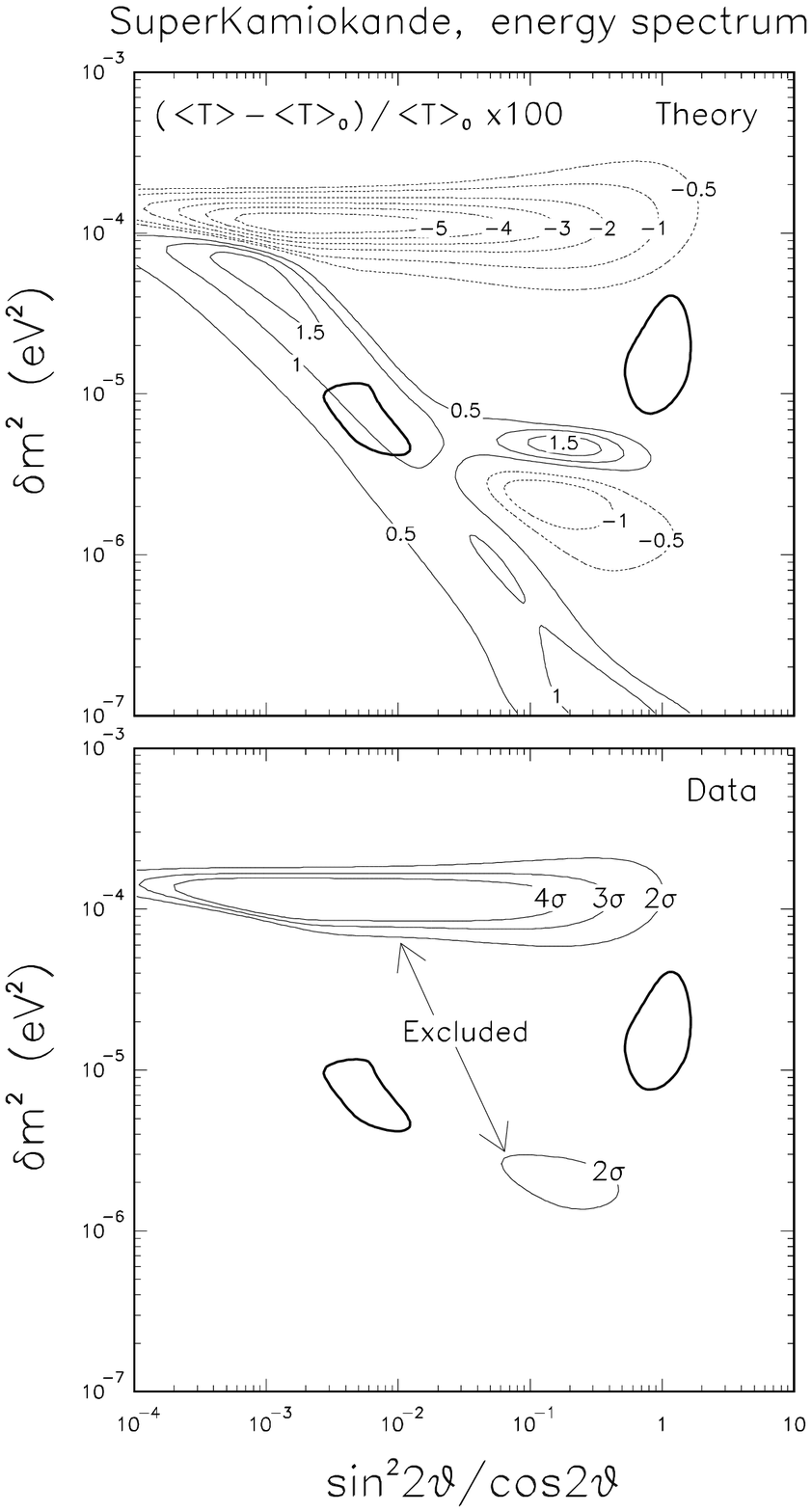}%
{FIG.~5. Constraints on the MSW solutions as derived by the measured value
	of the average electron kinetic energy $\langle T\rangle$. Upper 
	panel: Theoretical expectations for the fractional shift of 
	$\langle T\rangle$ from its standard (no oscillation) value 
	$\langle T \rangle_0$.  Lower panel: regions excluded at 2, 3, 
	and 4 standard deviations by the SuperKamiokande determination 
	of $\langle T\rangle$.}
%..........................................................................
\InsertFigure{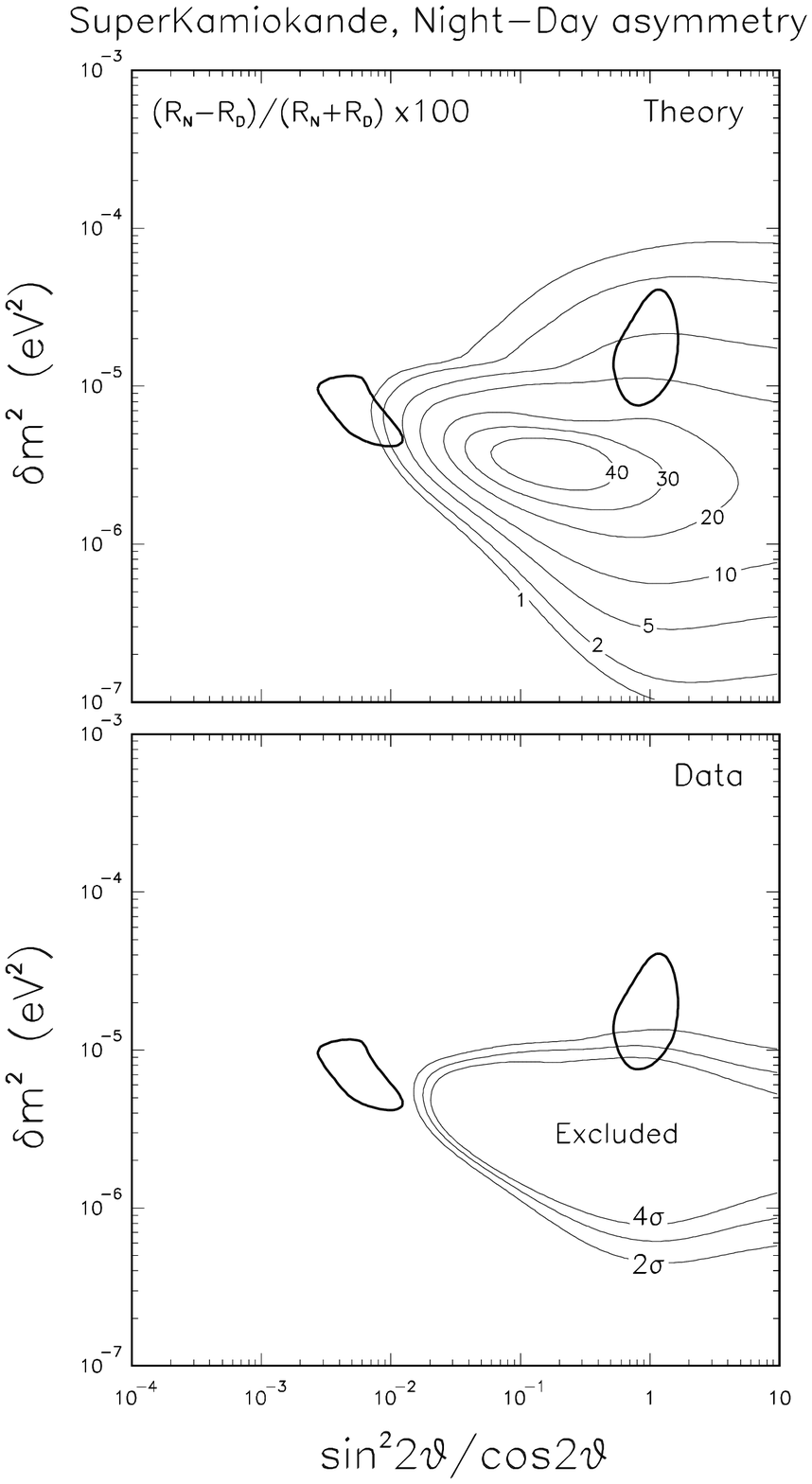}%
{FIG.~6. Constraints on the MSW solutions as derived by the measured
	value of the asymmetry of nighttime ($R_{\rm N}$) and daytime 
	($R_{\rm D}$) event rates. Upper panel: theoretical expectations 
	for \mbox{$(R_{\rm N}-R_{\rm D})/(R_{\rm N}+R_{\rm D})$}. Lower 
	panel:  Regions excluded at 2, 3, and 4 standard deviations by 
	the SuperKamiokande data.}
%..........................................................................
\InsertFigure{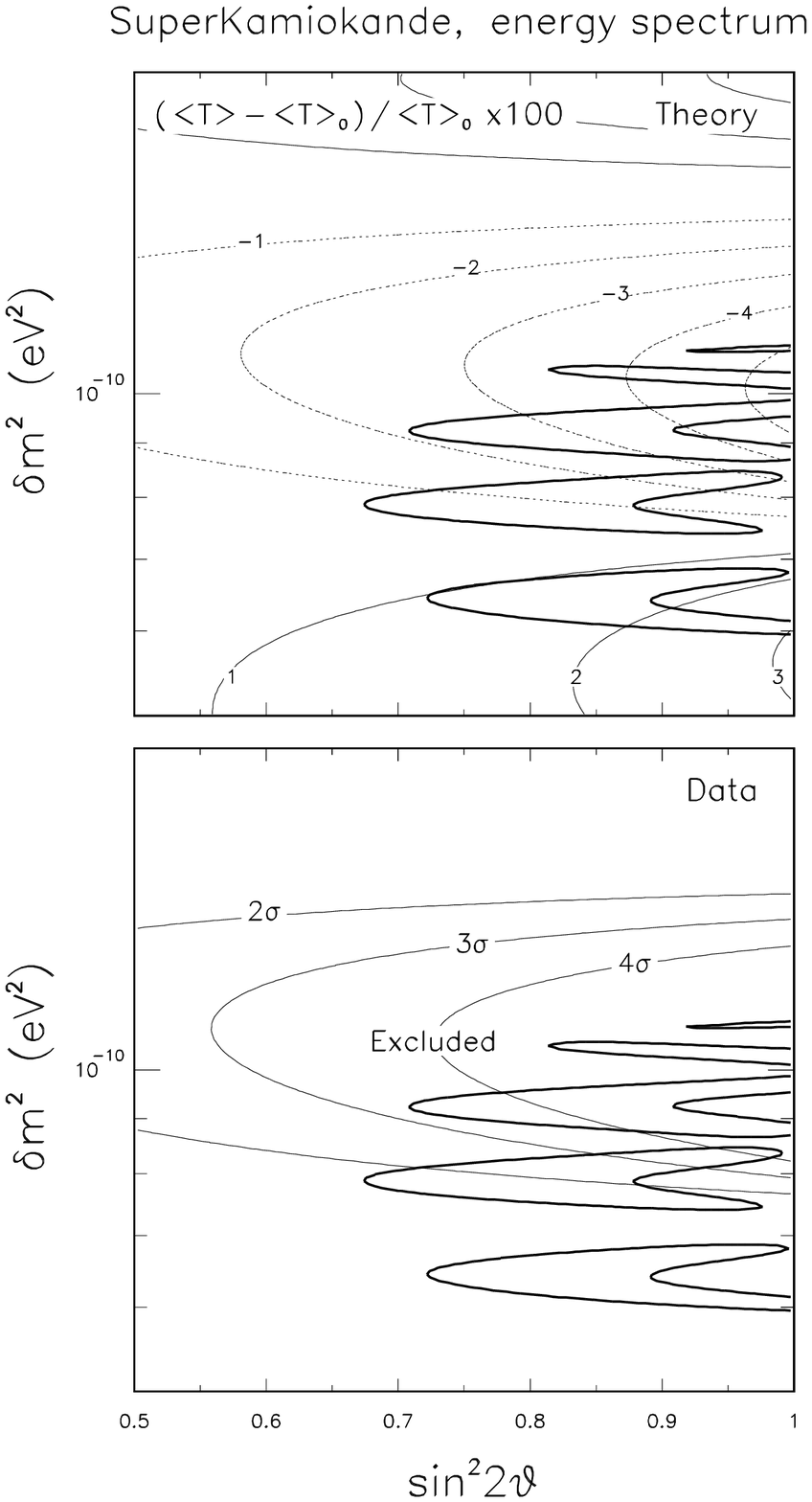}%
{FIG.~7. Constraints on the vacuum oscillation solutions as derived by the 
	measured value of the average electron kinetic energy 
	$\langle T\rangle$. Upper panel: Theoretical expectations for the 
	fractional shift of $\langle T\rangle$ from its standard 
	(no oscillation) value $\langle T \rangle_0$. Lower panel: Regions 
	excluded at 2, 3, and 4 standard deviations by the SuperKamiokande 
	determination of $\langle T\rangle$.}
%..........................................................................
\InsertFigure{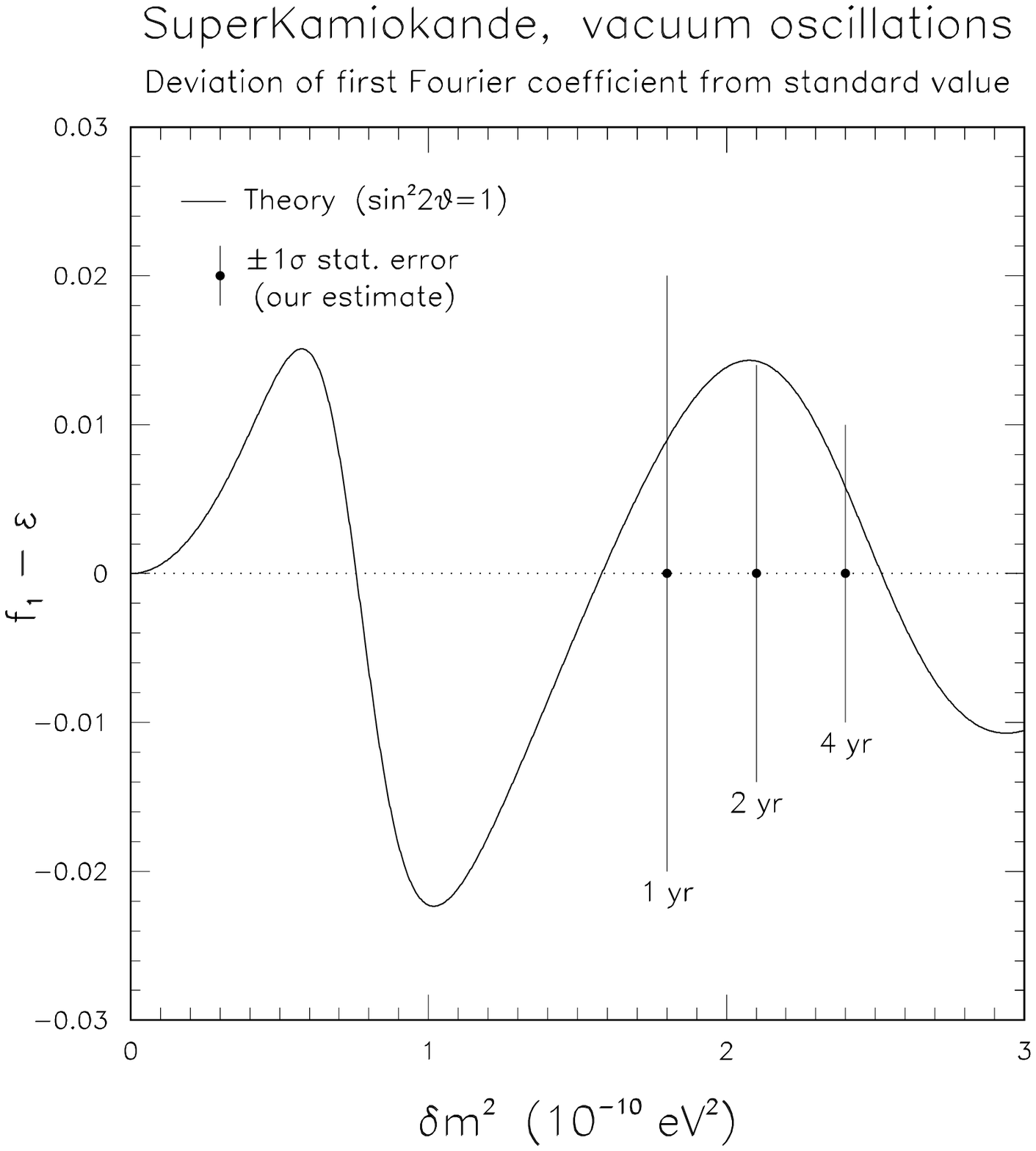}%
{FIG.~8. Fourier analysis of time variations induced by vacuum 
	oscillations. Solid curve: expected deviations of the first 
	Fourier coefficient $f_1$ from its standard value (equal to 
	Earth orbit eccentricity $\varepsilon$) for $\sin^22\theta=1$. 
	Error bars: Our estimate of the statistical errors of 
	SuperKamiokande (including background) after 1, 2, and 4 years 
	of data taking. See the text and \protect\cite{Four} for details.}
%..........................................................................

\end{document}